\documentclass[a4paper,11pt]{article}

\usepackage[utf8]{inputenc}
\usepackage{amsmath}
\usepackage{amssymb}
\usepackage{amsfonts}
\usepackage{mathrsfs}
\usepackage{graphicx}
\usepackage{booktabs,adjustbox}
\usepackage{caption}
\usepackage{slashed}
\usepackage{verbatim}
\usepackage{float}
\usepackage{setspace}
\usepackage{subfig}
\usepackage{jheppub}

\usepackage{color}
\usepackage{ulem}

\usepackage{bookmark}
\usepackage{wasysym}

\makeatletter
\newcommand{\thickhline}{%
	\noalign {\ifnum 0=`}\fi \hrule height 1pt
	\futurelet \reserved@a \@xhline
}
\makeatother

\allowdisplaybreaks

\title{\boldmath Power-aligned 2HDM: a correlative perspective on $(g-2)_{e,\mu}$}

\author[a]{Shao-Ping Li,}
\author[a,1]{ Xin-Qiang Li,\note{Corresponding author.}}
\author[a]{ Yuan-Yuan Li,}
\author[a]{Ya-Dong Yang}
\author[b]{and Xin Zhang}

\affiliation[a]{Institute of Particle Physics and Key Laboratory of Quark and Lepton Physics~(MOE), Central China Normal University, Wuhan, Hubei 430079, China}
\affiliation[b]{Faculty of Physics and Electronic Science, Hubei University, Wuhan, Hubei 430062, China}

\emailAdd{ShowpingLee@mails.ccnu.edu.cn}
\emailAdd{xqli@mail.ccnu.edu.cn}
\emailAdd{liyuanyuan@mails.ccnu.edu.cn}
\emailAdd{yangyd@mail.ccnu.edu.cn}
\emailAdd{xinzhang@hubu.edu.cn}

\abstract{With the hypothesis of minimal flavor violation, we find that there exists a power-aligned relation between the Yukawa couplings of the two scalar doublets in the two-Higgs-doublet model with Hermitian Yukawa matrices. Within such a power-aligned framework, it is found that a simultaneous explanation of the anomalies observed in the electron and muon anomalous magnetic moments can be reached with TeV-scale quasi-degenerate Higgs masses, and the resulting parameter space is also   phenomenologically safer under the B-physics, $Z$ and $\tau$ decay data, as well as the current LHC bounds. Furthermore, the flavor-universal power that enhances the charged-lepton Yukawa couplings prompts an interesting correlation between the two anomalies, which makes the model distinguishable from the (generalized) linearly aligned and the lepton-specific two-Higgs-doublet models that address the same anomalies but in a non-correlative manner, and hence testable by future precise measurements.}

\begin{document}
	\maketitle
	\flushbottom

\section{Introduction}

The minimally flavor-violating (MFV) hypothesis~\cite{DAmbrosio:2002vsn,Buras:2000dm} can be served as a guideline to construct the Yukawa interactions in effective field theories or explicit new physics (NP) models. In general, a global unitary symmetry $G_f$ that commutes with the Standard Model (SM) gauge group is assumed to be minimally broken by the Yukawa sector of the SM Lagrangian, and its restoration can be realized by promoting the Yukawa couplings to be auxiliary, non-dynamical fields, the so-called spurions, which transform non-trivially under the symmetry. For explicit implementations, however, the definition of the spurions depends crucially on the choice of the symmetry group. In the original $U(3)^5$ formulation~\cite{DAmbrosio:2002vsn}, the quark Yukawa sector breaks minimally the $U(3)^3$ symmetry, while the lepton Yukawa sector, with the neutrinos assumed to be massless, breaks the $U(3)^2$ symmetry. In recent development along the MFV criterion, the choice of the global symmetry is usually motivated by the low-energy phenomena. For instance, in refs.~\cite{Barbieri:2011ci,Barbieri:2011fc,Barbieri:2012uh}, instead of the largest $U(3)^3$ group in the quark sector~\cite{Gerard:1982mm,Chivukula:1987py}, a $U(2)^3$ flavor symmetry is considered, while in ref.~\cite{Li:2019xmi}, two of us have proposed a minimally broken $U(1)^3$ flavor symmetry in each of the quark and lepton sectors. These explicit constructions are all featured by the compelling MFV principle: with the SM Yukawa spurions defined, all the additional Yukawa interactions beyond the SM ones can be solely constructed in terms of the well-established fermion mass spectra and flavor mixings.  

The aligned two-Higgs-doublet model (A2HDM) postulated by Pich and Tuzon in ref.~\cite{Pich:2009sp} can also be derived from the MFV hypothesis~\cite{Buras:2010mh}. To this end, one need only keep the truncated spurions to the first order in the $U(3)^5$-breaking terms, rendering therefore the constructed NP Yukawa matrices to be linearly aligned with the SM ones, $Y_{\text{NP}}=\zeta\,Y_{\text{SM}}$, with $\zeta$ being arbitrary complex numbers~\cite{Pich:2009sp}. This ansatz makes the model free of flavor-changing neutral current (FCNC) at tree level, and the NP Yukawa effects depend crucially on the choices of the free flavor-universal parameters $\zeta$. In addition, the MFV setup helps to make the ansatz sufficiently stable beyond the tree level~\cite{Buras:2010mh}.

A prominent triumph of the A2HDM applications is that the $3.7\sigma$ discrepancy between the experimental measurement~\cite{Bennett:2006fi} and the SM prediction~\cite{Aoyama:2020ynm} of the muon anomalous magnetic moment, $a_\mu=(g-2)_\mu/2$, with $\Delta a_\mu\equiv a_\mu^{\rm exp}-a_\mu^{\rm SM}=(2.79\pm 0.76)\times 10^{-9}$,\footnote{For a comprehensive list of the SM predictions, we refer the readers to ref.~\cite{Aoyama:2020ynm} and the website, \url{https://muon-gm2-theory.illinois.edu/}, where the original references upon which the corresponding results are based could also be found. Note that a recent lattice-QCD calculation~\cite{Borsanyi:2020mff} of the leading-order hadronic vacuum polarization contribution to $(g-2)_\mu$ brings the SM prediction into agreement with the experimental data. However, this result is in tension with the $e^+e^-\to \text{hadrons}$ cross-section data and the global electroweak fits~\cite{Crivellin:2020zul}. For a different argument on this topic, see \textit{e.g.}, refs.~\cite{Malaescu:2020zuc,Keshavarzi:2020bfy}.} can be addressed by enhancing linearly the charged-lepton Yukawa couplings and ensuring large mass splittings among the additional Higgs bosons~\cite{Ilisie:2015tra,Han:2015yys,Cherchiglia:2016eui}. Interestingly, the Fermilab Muon $g-2$ experiment announced recently its first measurement of $a_\mu$~\cite{Abi:2021gix}, which is in full agreement with the previous measurement~\cite{Bennett:2006fi} and, once combined together, increases the significance of the discrepancy to the level of $4.2\sigma$, with~\cite{Abi:2021gix}
\begin{align}\label{eq:deltaamu}
	\Delta a_\mu\equiv a_\mu^{\rm exp}-a_\mu^{\rm SM}=(2.51\pm 0.59)\times 10^{-9}.
\end{align}
However, when confronted with the recently observed $2.4\sigma$ deviation between experiment~\cite{Hanneke:2008tm} and theory~\cite{Laporta:2017okg,Aoyama:2017uqe} in the electron anomalous magnetic moment, $a_e=(g-2)_e/2$, due to an improved measurement of the fine-structure constant~\cite{Parker:2018vye}, with
\begin{align}\label{eq:deltaae}
\Delta a_e\equiv a_e^{\rm exp}-a_e^{\rm SM}=-(8.7\pm 3.6)\times 10^{-13},
\end{align} 
the A2HDM can no longer provide an explanation simultaneously. This is because the linear alignment is flavor-universal and an explanation of $\Delta a_{\mu}$ fixes already the parameter space in which a totally positive NP effect on $\Delta a_{e}$ should arise. However, if one promotes the linearly aligned parameter $\zeta$ to a diagonal matrix, $\zeta=\text{diag}(\zeta_1,\zeta_2,\zeta_3)$, with three independent entries $\zeta_i$, a simultaneous explanation of both $\Delta a_{e}$ and $\Delta a_{\mu}$ anomalies can still be achieved~\cite{Jana:2020pxx,Botella:2020xzf}. Such a general A2HDM (gA2HDM)~\cite{Penuelas:2017ikk,Botella:2018gzy}, nevertheless, enlarges the degrees of freedom of the model parameters and makes the simultaneous explanation of the $\Delta a_{e,\mu}$ anomalies non-correlative. In addition to these phenomenological shortcomings, there exist also some theoretical subtleties when one follows the MFV spirit: if the generalized linear alignment is constructed in the fermion mass-eigenstate basis, the alignment condition cannot be guaranteed back in a general flavor basis, such as the fermion weak-eigenstate basis~\cite{Penuelas:2017ikk}; to maintain the alignment condition constructed in a general flavor basis, somewhat contrived commutation relations between the Yukawa matrices of the two scalar doublets or any other independent and linear combinations of these matrices are however required~\cite{Botella:2018gzy} (see also  refs.~\cite{Egana-Ugrinovic:2018znw,Egana-Ugrinovic:2019dqu} for a different setup in constructing the generalized Yukawa alignment in the framework of two-Higgs-doublet model (2HDM)). 

To circumvent these issues and realize a simultaneous and correlative explanation of the $\Delta a_{e,\mu}$ anomalies within the 2HDM framework, we firstly notice that, within the conventional MFV setup, various tantalizing Yukawa structures do not receive enough considerations. However, it is known that structural Yukawa matrices are intensively studied in the realm of deciphering the flavor puzzles of the SM and beyond (see, \textit{e.g.}, refs.~\cite{Fritzsch:1999ee,Altarelli:2010gt,Xing:2019vks,Feruglio:2019ktm} for comprehensive reviews). In this context, Hermitian Yukawa matrices are widely considered~\cite{Fritzsch:1979zq,Branco:1988iq}. For instance, the Hermitian matrices with four texture zeros in the quark~\cite{Fritzsch:1995nx,Mondragon:1998gy,Branco:1999nb,Fritzsch:2002ga,Grimus:2004hf,Xing:2015sva} as well as in the lepton sector with Dirac~\cite{Xing:2003zd,Ahuja:2007vh} or Majorana~\cite{Matsuda:2006xa,Branco:2007nn} neutrinos have been shown to predict well the observed patterns of quark and lepton mixings. Hermitian Yukawa matrices can also arise from more fundamental gauge theories, such as the \textit{manifestly} left-right symmetric models based on the gauge group $SU(2)_L\times SU(2)_R\times U(1)_{B-L}$~\cite{Mohapatra:1974gc,Senjanovic:1975rk,Mohapatra:1979ia,Mohapatra:1980yp}. When Hermitian Yukawa matrices meet the MFV guideline in an effective 2HDM framework, it can be demonstrated that the Hermitian Yukawa spurions, which recover the $U(3)^2$ rather than the largest $U(3)^5$ symmetry, allow the NP Yukawa matrices, $Y_{\rm NP}$, to be power-aligned with the SM ones, $Y_{\rm SM}$, with $Y_{\rm NP}=(Y_{\rm SM})^n$ ($n>0$ by assumption), as will be detailed in section~\ref{sec:mechanism}. In this way, the two Yukawa sectors will commute with each other, $[Y_{\rm SM},Y_{\rm NP}]=0$, guaranteeing therefore the absence of tree-level FCNC within the 2HDM framework.

In light of the proposed power-aligned Yukawa couplings, we will show further that a simultaneous and correlative explanation of the $\Delta a_{e,\mu}$ anomalies can be reached within the 2HDM framework. Explicitly, as the NP effect from the top-quark Yukawa coupling $y_t^{n_u}\simeq 0.99^{n_u}$ is insensitive to the choice of the power $n_u$, we find that, with the updated SM calculations of the mass differences $\Delta M_{d,s}$ of $B_{d,s}^0-\bar B_{d,s}^0$ mixing systems~\cite{Kirk:2017juj,King:2019lal,DiLuzio:2019jyq,Lenz:2019lvd}, the charged-Higgs mass is now pushed beyond a few TeV. With such an $\mathcal{O}(1)$~TeV-scale charged Higgs boson, the mass splittings among the new scalars should be reduced in order to comply with the electroweak precision tests and a set of experimental constraints from flavor observables~\cite{Haller:2018nnx}. Thus, large NP effects on $(g-2)_{e,\mu}$ cannot be produced due to the absence of large mass splittings among the scalars, contrary to the conclusions made in the alignment-based 2HDM~\cite{Ilisie:2015tra,Han:2015yys,Cherchiglia:2016eui,Jana:2020pxx} as well as the lepton-specific 2HDM (L2HDM)~\cite{Broggio:2014mna,Wang:2014sda,Abe:2015oca,Crivellin:2015hha,Chun:2016hzs,Wang:2018hnw,Han:2018znu}. For a sample of other recent suggestions made for the simultaneous explanation of the $\Delta a_{e,\mu}$ anomalies, we refer the readers to refs.~\cite{Davoudiasl:2018fbb,Crivellin:2018qmi,Liu:2018xkx,Endo:2019bcj,Bauer:2019gfk,Badziak:2019gaf,CarcamoHernandez:2019ydc,Hiller:2019mou,Cornella:2019uxs,Haba:2020gkr,Bigaran:2020jil,Calibbi:2020emz,Hati:2020fzp,Dutta:2020scq,Chen:2020tfr,Dorsner:2020aaz,Chun:2020uzw}.

More remarkably, large mass splittings and linear alignment that are required to address the $\Delta a_\mu$ anomaly at $1\sigma$ level would also cause large NP effects on the lepton-flavor universality tests in $Z$ and $\tau$ decays~\cite{Abe:2015oca,Chun:2016hzs}. Within the \textit{power-aligned} 2HDM (pA2HDM) proposed here, instead, the resolution of $\Delta a_{e,\mu}$ can alleviate the constraints from these leptonic precision observables, because, on the one hand, the new Higgs mass spectrum is now quasi-degenerate, rendering some cancellations among the quantum corrections, and, on the other hand, the power enhancement in the electron and muon Yukawa entries necessary for the $\Delta a_{e,\mu}$ explanation increases the tau Yukawa coupling at a speed slower than that in the linear alignment. Furthermore, being distinguishable from the gA2HDM that addresses the $\Delta a_{e,\mu}$ anomalies in a non-correlative manner~\cite{Jana:2020pxx,Botella:2020xzf}, the pA2HDM predicts an interesting correlation between the two anomalies, which can be therefore tested by future precise measurements~\cite{Grange:2015fou,Abe:2019thb,Abbiendi:2016xup}. 

This paper is organized as follows. In section~\ref{sec:mechanism}, with the MFV hypothesis, we propose the pA2HDM with an additional Hermiticity condition of the Yukawa matrices. In section~\ref{sec:g-2}, we firstly present the experimental constraints from B-physics observables that push the charged-Higgs mass up to a few TeV, and then analyze the Higgs mass spectrum with a $Z_2$-symmetric scalar potential. We will show that these preconditions render a simultaneous and correlative explanation of the $\Delta a_{e,\mu}$ anomalies with TeV-scale quasi-degenerate Higgs masses, and discuss such an explanation under the constraints from $Z$ and $\tau$ decay data, as well as the current LHC bounds. Our conclusions are finally made in section~\ref{sec:con}.

\section{Power-aligned 2HDM: Hermiticity meets MFV}
\label{sec:mechanism}

Within a generic 2HDM framework~\cite{Branco:2011iw}, the Yukawa interactions at the electroweak gauge symmetric regime can be attributed to the only source that violates explicitly the flavor symmetry groups $SU(3)_{Q}\times SU(3)_{u}\times SU(3)_{d}$ in the quark and $SU(3)_{E}\times SU(3)_{\ell}\times SU(3)_\nu$ in the lepton sector\footnote{Note that the Abelian $U(1)$ subgroups of $U(3)^5$ can be identified to be associated with some conserved charges, such as the baryon and lepton numbers~\cite{DAmbrosio:2002vsn,Chivukula:1987py}, and we will not consider these $U(1)$ factors hereafter.}, where $Q$ and $E$ denote respectively the quark and lepton $SU(2)_L$ doublets, while $u$, $d$ and $\ell$ are the right-handed fermion $SU(2)_L$ singlets. Here we have simply embedded three right-handed Dirac neutrinos into the SM, with the neutrino masses generated via the Higgs mechanism. In accordance with the MFV hypothesis~\cite{DAmbrosio:2002vsn,Buras:2000dm}, the corresponding SM Yukawa couplings are promoted to be spurions with the following non-trivial transformation properties under the symmetry~\cite{DAmbrosio:2002vsn}:
\begin{align}
Y_u\sim (3_Q,\bar{3}_u,1_d),\quad Y_d\sim (3_Q,1_u,\bar 3_d),\quad 
Y_\ell \sim (3_E,\bar{3}_\ell,1_\nu),\quad Y_\nu\sim (3_E,1_\ell,\bar{3}_\nu).
\end{align} 
Then, any additional interactions invariant under the electroweak gauge group should be built up with these definite spurions as well as the SM fields. In this way, the unitary flavor symmetry can be recovered even in the Yukawa Lagrangian at tree level. The compelling consequence of the MFV criterion is that all the NP effects can be well described in terms of the known fermion mass spectra and flavor mixings~\cite{DAmbrosio:2002vsn}. However, the unitary field transformations do not correspond to a realistic flavor symmetry, and hence the absence of FCNC is not protected from renormalization-group (RG) running effects~\cite{Ferreira:2010xe}. Nevertheless, it has been shown that the RG-induced FCNC effects are small and still comply with the current experimental observations~\cite{Penuelas:2017ikk,Braeuninger:2010td,Gori:2017qwg,Jung:2010ik,Li:2014fea}.

As mentioned in the Introduction, the Hermitian Yukawa matrices could originate from some fundamental theories. Let us take here the manifestly left-right symmetric model~\cite{Mohapatra:1974gc,Senjanovic:1975rk,Mohapatra:1979ia,Mohapatra:1980yp} as an illustrating example, to show that the unitary flavor transformations are reduced to an $SU(3)_q$ in the quark and an $SU(3)_l$ in the lepton sector, when the Hermitian Yukawa matrices are predicted by some flavor and/or gauge symmetry. For this purpose, it suffices to consider the following $SU(2)_L\times SU(2)_R\times U(1)_{B-L}$-invariant Yukawa interactions:
\begin{align}\label{Yukawa in LR}
\mathcal{L}^{LR}_Y \supset Y_a  \bar F_L \Phi_a F_R 
+{\rm H.c.},
\end{align}
where $F_{L,R}$ are the fermion $SU(2)_{L,R}$ doublets, $\Phi_a$ ($a=1,2,\cdots$) the Higgs bi-doublets with representation $(2,2,0)$, and $Y_a$ the Yukawa matrices associated with $\Phi_a$. Besides the gauge symmetry, there exists a discrete left-right parity symmetry under which the interchanges $F_L\leftrightarrow F_R$ and $\Phi_a\leftrightarrow \Phi_a^\dagger$ also keep the Lagrangian invariant. Then, it follows from eq.~\eqref{Yukawa in LR} that the Yukawa matrices $Y_a$ are Hermitian, $Y_a=Y_a^{\dagger}$. In this case, the unitary field transformations that commute with the gauge as well as the left-right parity symmetry should be reduced to a single $SU(3)_{q}\times SU(3)_l$ group, such that under the transformation
\begin{align}
\bar F_L Y_a F_R \to (\bar F_L V_F^\dagger) (V_F Y_a V_F^\dagger) (V_F F_R),
\end{align}
the Hermiticity of the Yukawa matrices in the new basis, $V_F Y_aV_F^\dagger$, can be maintained. In the rest of this work, we will refrain from discussing detailed model buildings, but focus only on the effective 2HDM framework that could stem from some fundamental theory in which the Hermitian Yukawa matrices are predicted. One of such possibilities is that the effective 2HDM has an origin from the manifestly left-right symmetric models with multi-Higgs bi-doublets (see, \textit{e.g.}, ref.~\cite{Iguro:2018oou}). The only required precondition is that, if the Yukawa matrices are Hermitian in the underlying theory equipped with the MFV principle, they should be constructed in terms of the minimal spurions, and this observation is inherited down to the effective 2HDM framework. Then, in the non-decoupled 2HDM regime, one should be able to discern the $SU(3)_{q}\times SU(3)_l$-invariant pattern via the NP effects exerted on the low-energy observables. 

Given that the Yukawa spurions now transform under $SU(3)_{q}\times SU(3)_l$ as
\begin{align}
Y_u\sim (3_q,\bar{3}_q),\quad Y_d \sim (3_q,\bar{3}_q),\quad 
Y_\ell \sim (3_l,\bar{3}_l),\quad Y_\nu \sim (3_l,\bar{3}_l),
\end{align} 
all the Yukawa interactions beyond the SM ones that are in accordance with the MFV criterion should be constructed in terms of the following linear combinations of the spurions:
\begin{align}\label{spurion products}
Y_U,\quad Y_D,\quad (Y_U^2)Y_U,\quad (Y_D^2)Y_D,\quad (Y_U Y_D) Y_U, \quad (Y_D Y_U) Y_D,\quad \cdots,
\end{align}
where the subscripts $U$ and $D$ denote the up- and down-type fermions, respectively. Starting with eq.~\eqref{spurion products}, let us now consider what kind of simple Yukawa alignment can be constructed in a MFV manner, from the following most general Yukawa Lagrangian in a generic 2HDM framework~\cite{Branco:2011iw}:
\begin{align}\label{glag}
-\mathcal L_Y &= \bar{Q}_L\left(Y_1^u \tilde{H}_1+Y_2^u \tilde{H}_2\right) u_R+ \bar{Q}_L\left(Y_1^d H_1 + Y_2^d H_2\right) d_R
\nonumber \\[0.2cm]
& + \bar{E}_L\left(Y_1^\ell H_1 +Y_2^\ell H_2\right) e_R + \bar{E}_L\left(Y_1^{\nu}\tilde{H}_1+Y_2^{\nu} \tilde{H}_2\right)\nu_R + {\rm H.c.},
\end{align}
where the two Higgs doublets are parametrized, respectively, as
\begin{eqnarray}\label{Higgs param}
H_1 =\left(
\begin{array}{c}
G^+\\
\frac{\phi_1+iG^0}{\sqrt{2}}
\end{array}
\right),\qquad
H_2=\left(
\begin{array}{c}
H^+\\
\frac{\phi_2+iA}{\sqrt{2}}
\end{array}
\right).
\label{HiggsBasis}
\end{eqnarray}

Let us firstly recall that, with the spirit of MFV hypothesis, the Yukawa interactions are the only source that breaks certain symmetries satisfied by other parts of the whole Lagrangian. Then, it is natural to expect that some symmetry preserved by the Higgs sector (kinetic terms and scalar potential) would also be broken by the Yukawa part. On the other hand, the FCNC effects mediated by the SM Higgs boson, which arise from the mixing of the neutral scalars $\phi_{1,2}$, have already been severely constrained by the observed Higgs signals at the Large Hadron Collider (LHC) (see \textit{e.g.}, ref.~\cite{Haller:2018nnx} for an updated review). Based on these observations, we are motivated to consider an exactly $Z_2$-symmetric scalar potential in which the parities of the two Higgs doublets are defined, respectively, by $\mathcal{Z}_2(H_1)=1$ and $\mathcal{Z}_2(H_2)=-1$. In this case, the vacuum expectation value of $\phi_2$ should vanish, $\langle\phi_2\rangle=0$, in order to preserve the $Z_2$ symmetry in the scalar potential, while $\langle\phi_1\rangle=v\simeq 246$~GeV is responsible for generating the fermion masses. Then, $G^{\pm,0}$ become the Goldstone bosons, and the excitation of $\phi_1$, $\phi_1=v+h$, boils down to the SM Higgs boson (a similar construction can be found, \textit{e.g.}, in ref.~\cite{Li:2019xmi}). As will be discussed in section~\ref{Higgs spectrum}, the $Z_2$ symmetry manifested in the scalar potential can also protect the quasi-degenerate Higgs mass spectrum from large radiative corrections. 

In the A2HDM~\cite{Pich:2009sp}, the simplest Yukawa construction from the infinite set of spurions given by eq.~\eqref{spurion products} is adopted, with the linear alignment realized by~\cite{Buras:2010mh}
\begin{align}
Y_2^f=\zeta_f Y_1^f,
\end{align}
where $\zeta_f$ are flavor-universal proportionality parameters. In the gA2HDM~\cite{Jana:2020pxx,Botella:2020xzf,Penuelas:2017ikk,Botella:2018gzy}, on the other hand, the parameters $\zeta_f$ are promoted to be diagonal matrices with three independent entries in the mass-eigenstate basis. It is readily to see that the FCNC effects are absent at tree level in both cases. Specific to our case, since the Yukawa matrices are now presumed to be Hermitian, there exists another simple alignment construction from eq.~\eqref{spurion products}, with
\begin{align}\label{PL in general basis}
Y_2=Y_1^n.
\end{align}
It is trivial to see that $[Y_1,Y_2]=0$ for positive integers $n\in \boldsymbol{Z^+}$. For rational fractions in the interval $0<n<1$, with $m=1/n\in \boldsymbol{Z^+}$, on the other hand, one can see that
\begin{align}\label{commutation for 1/n}
[Y_1,Y_2]=[Y_1,Y_1^n]=[\tilde{Y}^m,\tilde{Y}]=0,
\end{align} 
where $\tilde{Y}\equiv Y_1^n$. Together with these two observations, it is straightforward to check that $[Y_1,Y_2]=0$ also holds for other positive rational fractions $n$, with $1/n$ being not integers. Thus, with the prescription given by eq.~\eqref{PL in general basis}, the two Hermitian Yukawa matrices $Y_{1,2}$ always commute with each other and can be, therefore, diagonalized simultaneously. For the present work, nevertheless, it suffices to consider the case where $n\in \boldsymbol{Z^+}$ or $1/n\in \boldsymbol{Z^+}$.

After spontaneous gauge symmetry breaking, the Yukawa interactions in the fermion mass-eigenstate basis can be formally written as  
\begin{align}\label{lag}
-\mathcal{L}_Y &= \bar{Q}_L V^\dagger \left(\hat{Y}_u \tilde{H}_1+\mathcal{Y}_u \tilde{H}_2\right) u_R + \bar{Q}_L \left( \hat{Y}_d H_1+\mathcal{Y}_d H_2\right) d_R \nonumber \\[0.1cm]
&+ \bar{E}_L U\left(\hat{Y}_{\nu}\tilde{H}_1+\mathcal{Y}_{\nu} \tilde{H}_2\right)\nu_R + \bar{E}_L\left( \hat{Y}_\ell H_1+\mathcal{Y}_\ell H_2 \right)  e_R+ {\rm H.c.},
\end{align}
where $Q_L\equiv (V^\dagger u_L, d_L)^T$ and $E_L\equiv (U\nu_L, e_L)^T$, with $V=V_u V_d^\dagger$ and $U=V_\ell V_\nu^\dagger$ being the Cabibbo-Kobayashi-Maskawa (CKM)~\cite{Cabibbo:1963yz,Kobayashi:1973fv} and Pontecorvo-Maki-Nakagawa-Sakata (PMNS)~\cite{Pontecorvo:1957cp,Maki:1962mu} matrices, respectively. $\hat{Y}_f$ are the SM diagonal Yukawa matrices, $\hat{Y}_f=\text{diag}(y_{f_1},y_{f_2},y_{f_3})$, with $y_{f_i}$ being real and positive, while $\mathcal{Y}_f$ encode all the NP Yukawa interactions. Here our convention for rotation from the flavor ($f^\prime$) to the mass ($f$) eigenstates is defined as $f^\prime=V_f^{\dagger}f$. The relations between Hermitian ($Y_{1,2}^f$), diagonal ($\hat{Y}_f$) and NP ($\mathcal{Y}_f$) Yukawa matrices are then given by
\begin{align}\label{diagonalization}
Y_1^{f}&=V^{\dagger}_f\hat{Y}_f V_f,
\nonumber \\[0.2cm]
Y_2^{f}&=(Y_1^{f})^{n_f}=(V^{\dagger}_f \hat{Y}_fV_f)^{n_f}\equiv (V^{\dagger}_f \mathcal{Y}_fV_f).
\end{align}
It is trivial to see that, for $n_f\in \boldsymbol{Z^+}$, we have the power alignment
\begin{align}\label{PL in mass basis} 
\mathcal{Y}_f=(\hat{Y}_f)^{n_f}.
\end{align}
For $1/n_f\in \boldsymbol{Z^+}$, on the other hand, following eq.~\eqref{commutation for 1/n}, we can rewrite the second line in eq.~\eqref{diagonalization} as
\begin{align}\label{PL in mass basis2} 
(Y_2^f )^{1/n_f}\equiv (V^{\dagger}_f \mathcal{Y}_fV_f)^{1/n_f}=(V^{\dagger}_f \hat{Y}_fV_f),
\end{align}
where the second relation is obtained by the power condition in flavor basis (see eq.~\eqref{PL in general basis}) and it leads to $(\mathcal{Y}_f)^{1/n_f}=\hat{Y}_f$. In this case, eq.~\eqref{PL in mass basis} can be derived up to a phase difference $\theta =2k n_f \pi$, with natural numbers $k=0,1,\cdots,1/n_f-1$. Here $\theta$ is the phase in each diagonal entry of $\mathcal{Y}_f$, $\mathcal{Y}_{f_i}=e^{i \theta_i} \tilde{y}_{f_i}$, with $\tilde{y}_{f_i}=(y_{f_i})^{n_f}$. 

Therefore, the entries in $\mathcal{Y}_f$ are in general complex if $1/n_f\in \boldsymbol{Z^+}$, but always positive definite if $n_f\in \boldsymbol{Z^+}$. However, we will not consider non-trivial phases in the diagonal entries, which would confront tight constraints from the electric dipole moments of elementary particles or systems~\cite{Jung:2013hka,Abe:2013qla,Inoue:2014nva,Cheung:2014oaa,Kanemura:2020ibp,Altmannshofer:2020shb} (see, \textit{e.g.}, refs.~\cite{Bernreuther:1990jx,Pospelov:2005pr,Engel:2013lsa,Chupp:2017rkp} for reviews on this subject), but focus on the CP-conserving case with $\theta_i=0,\pi$. In particular, it will be shown in the next section that the situation with $\theta_e=0$ and $\theta_\mu=\pi$ allows a simultaneous and correlative explanation of the $\Delta a_{e,\mu}$ anomalies, which, nevertheless, requires an even integer for $1/n_\ell$. It will also be shown that the parameter region allowed by the simultaneous explanation of $\Delta a_{e,\mu}$ anomalies \textit{does} intriguingly prompt an even integer for $1/n_\ell$.

Finally, it should be mentioned again that, as a common feature in the MFV-based 2HDM setup, the truncated spurions constructed at the high-energy scale are not protected from the RG running effect, and thus there are RG-induced FCNCs~\cite{Ferreira:2010xe}. However, as illustrated already in refs.~\cite{Penuelas:2017ikk,Braeuninger:2010td,Gori:2017qwg}, the violation of alignment due to the RG-induced FCNCs is small. It is, therefore, a good approximation to take the \textit{classical} Yukawa patterns to analyze the NP effects on low-energy observables. 

\section{Tying the explanations of $\boldsymbol{\Delta a_{e,\mu}}$ anomalies}
\label{sec:g-2}

The NP effect on $(g-2)_\mu$ in the L2HDM has been studied intensively in refs.~\cite{Broggio:2014mna,Wang:2014sda,Abe:2015oca,Chun:2016hzs,Wang:2018hnw,Han:2018znu}. It was found that a linear enhancement of the muon Yukawa coupling and a large mass splitting between the scalar and pseudoscalar Higgs bosons cannot explain the $\Delta a_\mu$ anomaly at $1\sigma$ level, because such a setup would violate the severe lepton-flavor universality tests in $Z$ and $\tau$ decays~\cite{Abe:2015oca,Chun:2016hzs}. The situation can be somewhat alleviated in the A2HDM~\cite{Ilisie:2015tra,Han:2015yys,Cherchiglia:2016eui}, since a significant contribution appears in the two-loop Barr-Zee diagrams involving the top-quark propagator in the loop, though a large mass splitting is still required to be at work. Being different from these setups, we will show in this section step by step that, in the pA2HDM, the current constraints from B-physics observables already push the charged-Higgs mass up to a few TeV, and the well-known electroweak precision tests, together with a small scalar-potential parameter $\lambda_5$, infer a quasi-degenerate Higgs mass spectrum, with $M_H\simeq M_A\simeq M_{H^+}$ at the TeV scale. Under these combined constraints, and further due to a flavor-universal power $n_f$, a simultaneous and correlative explanation of the $\Delta a_{e,\mu}$ anomalies can be achieved within $1\sigma$ level, and the resulting parameter space also complies with the $Z$ and $\tau$ decay data as well as the current LHC bounds. 

\subsection{Pushing up the charged-Higgs mass by B-physics observables}

In the pA2HDM, as the NP top-quark Yukawa coupling is given by $\vert \mathcal{Y}_{u,3}\vert=y_t^{n_u}\simeq0.99^{n_u}$, with the input $m_t=172.76$~GeV~\cite{Zyla:2020zbs}, the resulting contributions involving this coupling are insensitive to the choice of the power $n_u$. Given that the down-type NP Yukawa couplings are already constrained severely by the B-physics observables (see, \textit{e.g.}, ref.~\cite{Li:2018aov} and references therein), we will consider in this sector the power-suppressed effect with $n_d>1$. Nevertheless, even with such a choice, the NP top-quark Yukawa coupling can still give a large effect on some B-physics observables. In this context, the branching ratio $\mathcal{B}_{s\gamma}$ of the inclusive radiative $\bar B \to X_s \gamma$ decay and the mass differences $\Delta M_{d,s}$ of the $B_{d,s}^0-\bar B_{d,s}^0$ mixing systems can receive large $H^+$-$t$ loop corrections, and would depend only on the charged-Higgs mass $M_{H^+}$ in our approximation. 

Concentrating on the regime with $n_d>1$, we now follow ref.~\cite{Li:2018aov} to calculate the $H^+$-$t$ loop corrections to the branching ratio $\mathcal{B}_{s\gamma}$ and the mass differences $\Delta M_{d,s}$. The recently updated SM prediction, $\mathcal{B}_{s\gamma}^{\rm SM}=(3.40\pm 0.17)\times 10^{-4}$~\cite{Misiak:2020vlo}, and the experimental world average, $\mathcal{B}_{s\gamma}^{\rm exp}=(3.32\pm 0.15)\times 10^{-4}$~\cite{Zyla:2020zbs,Amhis:2019ckw}, both of which are given with a photon-energy cutoff $E_\gamma>1.6$~GeV, will be used. For the updated SM calculations of $\Delta M_{d,s}$, we will adopt the 2019 results presented in refs.~\cite{DiLuzio:2019jyq,Lenz:2019lvd}, which used a weighted average for the hadronic matrix elements obtained from lattice simulations~\cite{Aoki:2019cca,Dowdall:2019bea,Boyle:2018knm,Bazavov:2016nty} and sum rules~\cite{Kirk:2017juj,King:2019lal,Grozin:2016uqy}, resulting in
\begin{align}
\Delta M_d^{\rm SM-2019}=(0.533^{+0.022}_{-0.036})\text{ps}^{-1}, \qquad \Delta M_s^{\rm SM-2019}=(18.4^{+0.7}_{-1.2})\text{ps}^{-1}.
\end{align}
The updated SM predictions are now compatible with the current world averages~\cite{Zyla:2020zbs,Amhis:2019ckw},
\begin{align}
\Delta M_d^{\rm exp}=\left(0.5065\pm 0.0019 \right)\text{ps}^{-1},\qquad 
\Delta M_s^{\rm exp}=\left(17.757\pm 0.021 \right)\text{ps}^{-1},
\end{align}
within $1\sigma$ error bar. The good agreement between theory and experiment for both $\mathcal{B}_{s\gamma}$ and $\Delta M_{d,s}$ will put stringent constraints on the charged-Higgs mass $M_{H^+}$. In addition, the constraint from $Z\to\bar b b$ decay obtained in ref.~\cite{Jung:2010ik}, once translated to the case in the pA2HDM framework, will result in
\begin{align}
(n_u-1)>\ln\left(0.0024\,M_{H^+}/\text{GeV}+0.72\right)/\ln\left(\sqrt{2}m_t/v\right).
\end{align}

\begin{figure}[ht]
	\centering	
	\includegraphics[width=0.68\textwidth]{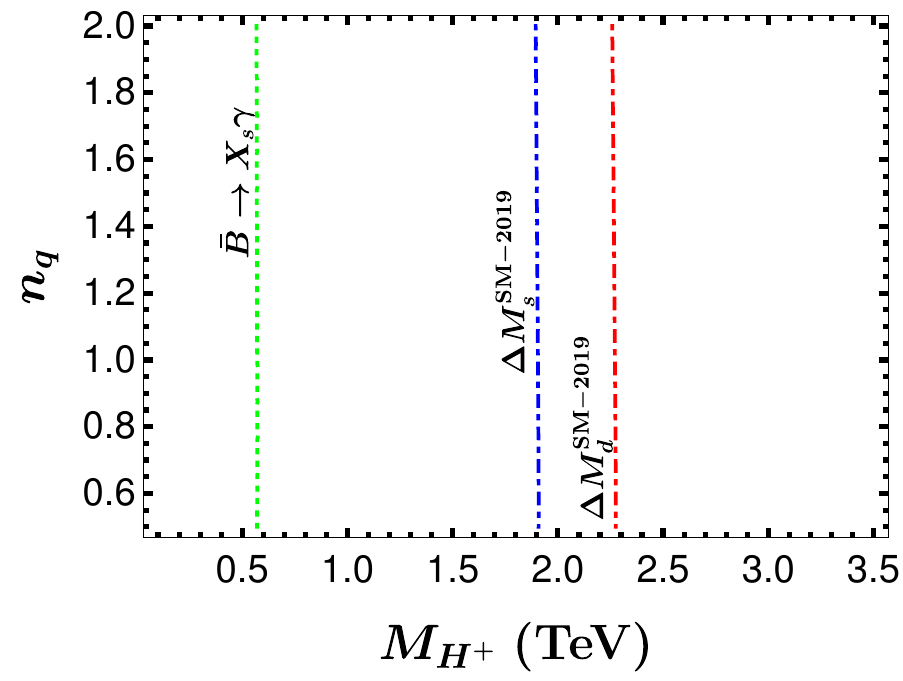}
	\caption{\label{Bconstraints} Lower bounds on the charged-Higgs mass $M_{H^+}$ from the $\bar B \to X_s \gamma$ branching ratio and the mass differences $\Delta M_{d,s}$. Here the $2\sigma$ ranges of the ratios $\Delta M_{d,s}^{\rm exp}/\Delta M_{d,s}^{\rm SM}$, with the 2019 updated $\Delta M_{d,s}^{\rm SM}$ from refs.~\cite{DiLuzio:2019jyq,Lenz:2019lvd}, are used as constraints.}
\end{figure}

Applying the experimental constraints mentioned above, we obtain the lower bounds on the charged-Higgs mass $M_{H^+}$, as shown in figure~\ref{Bconstraints}. The constraint from $Z\to \bar b b$ decay is found to be much weaker and does not impose any further restriction on the parameter region shown in the figure. The constraint from $\bar B \to X_s \gamma$ branching ratio is only sensitive to the charged-Higgs mass below $600$~GeV, while the mass differences $\Delta M_{d,s}$ put the most stringent bound on $M_{H^+}$. Such a restriction is, however, quite significant with respect to the SM predictions for $\Delta M_{d,s}$. In particular, the 2019 updated prediction, $\Delta M_{d}^{\rm SM-2019}$, pushes already the charged-Higgs mass beyond $2$~TeV. 

\subsection{Quasi-degenerate Higgs mass spectrum}
\label{Higgs spectrum}

For our purpose, we will consider the scalar potential with a $Z_2$ symmetry, which reads~\cite{Branco:2011iw}
\begin{align}\label{HiggsBasis potential}
V(H_1,H_2)&=M_{11}^2 H_1^\dagger H_1+M_{22}^2 H_2^\dagger H_2
+\frac{\lambda_1}{2}(H_1^\dagger H_1)^2+\frac{\lambda_2}{2}(H_2^\dagger H_2)^2+\lambda_3(H_1^\dagger H_1)(H_2^\dagger H_2)
 \nonumber \\[0.2cm]
& +\lambda_4 (H_1^\dagger H_2)(H_2^\dagger H_1)+\frac{\lambda_5}{2}
 \left[(H_1^\dagger H_2)^2+{\rm H.c.}\right],
\end{align}
where the two Higgs doublets $H_{1,2}$ are given by eq.~\eqref{Higgs param}.
In addition, we have assumed that the scalar potential is CP-conserving, which implies that the coupling $\lambda_5$ is real. The $Z_2$ symmetry is only broken by the Yukawa interactions (see eq.~\eqref{lag}), which is in line with the spirit of MFV hypothesis. The corresponding Higgs mass spectrum is given by
\begin{align}\label{Higgs_spectrum2}
m_h^2& = \lambda_1 v^2,\qquad m_H^2 = m_A^2+\lambda_5 v^2,
\nonumber \\[0.2cm]
m_A^2 &= m_{H^+}^2+\frac{\lambda_4-\lambda_5}{2} v^2,
\qquad m_{H^+}^2=M_{22}^2+\frac{v^2}{2}\lambda_3.
\end{align}

Before demonstrating that the mass spectrum given by eq.~\eqref{Higgs_spectrum2} can be quasi-degenerate, let us provide here some comments on the choice of a $Z_2$-symmetric scalar potential specified by eq.~\eqref{HiggsBasis potential}. Firstly, we should note that there might be a subtlety concerning the vacuum structure $\langle H_2 \rangle=(0, 0)^T$ in eq.~\eqref{Higgs param}. In a generic 2HDM framework, such a vacuum structure could result from a specific unitary transformation from a more general scalar basis $(\Phi_1,\Phi_2)$ with $\langle \Phi_{1,2}^0 \rangle\neq 0$. If this is indeed the case, the constructed Yukawa interactions, especially when their Yukawa matrices satisfy the power-aligned relations given by eq.~\eqref{PL in general basis}, would boil down to a particular choice of the scalar basis, making the power parameters $n_f$ being basis dependent in a very non-trivial way. Then we need further exploit the basis dependence of the Yukawa alignment. However, if the scalar potential considered has a $Z_2$ symmetry with definite $Z_2$ parities for the two Higgs doublets $H_{1,2}$, and this imposed symmetry is also maintained even after gauge symmetry breaking, the vacuum structure $\langle H_2^0 \rangle=0$ would be a consequence of the maintenance of $Z_2$ symmetry rather than of the non-trivial scalar basis transformation. On the other hand, it should be mentioned that the $Z_2$-maintained vacuum structure also allows multi-TeV Higgs bosons without spoiling the requirements of perturbative unitarity of high-energy $2\to 2$ scalar scatterings as well as of perturbativity of the quartic couplings (see \textit{e.g.}, refs.~\cite{Horejsi:2005da,Biswas:2014uba}), since the mass parameter $M_{22}$ in the scalar potential now does not participate in the minimization conditions, and is independent of the constrained quartic couplings (see ref.~\cite{Nebot:2020niz} for a recent discussion). Therefore, additional TeV-scale Higgs bosons can be realized in our framework, even with a $Z_2$-symmetric scalar potential without invoking the soft-breaking term $M_{12}^2 H_1^\dagger H_2+\text{H.c.}$, which is contrary to the observation made, \textit{e.g.}, in refs.~\cite{Horejsi:2005da,Biswas:2014uba,Gunion:2002zf}.

It is already known that, with a quasi-degenerate mass relation $M_{H^{+}}\simeq M_A$ (protected by a custodial symmetry~\cite{Haber:2010bw,Toussaint:1978zm}) or $M_{H^{+}}\simeq M_H$ (protected by a twisted custodial symmetry~\cite{Gerard:2007kn}), the NP effects on electroweak precision tests and a set of flavor-physics observables are guaranteed to be suppressed due to some delicate cancellations. Furthermore, the heavier the Higgs bosons, the smaller will be the mass splittings among them~\cite{Haller:2018nnx}. Thus, when the charged-Higgs mass $M_{H^+}$ is pushed up to a few TeV, either $M_A$ or $M_H$ should also reside at that scale. In fact, the mass quasi-degeneracy $M_{A}\simeq M_H$ can also be obtained if the coupling $\lambda_5$ is small, as can be seen from eq.~\eqref{Higgs_spectrum2}. Since the term associated with $\lambda_5$ is the only breaking source of a global $U(1)$ symmetry in the scalar potential (see eq.~\eqref{HiggsBasis potential}), the one-loop corrections to $M_{H,A}$ from tri-linear scalar couplings would also be proportional to the small coupling $\lambda_5$. Certainly, the loop corrections to $M_{H,A}$ can also arise from the Yukawa interactions, with the dominant contribution being due to the top-quark loop. This will result in the mass difference $\delta M_{HA}=\vert M_H-M_A\vert$ between the two scalars, with
\begin{align}
\delta M_{HA}\simeq \frac{3\mathcal{Y}_u^2 M_S}{8\pi^2}\left[x \ln x -x (1-4x)^{1/2}\,\ln\left(\frac{1-(1-4x)^{1/2}}{2x}-1\right)-2x\right],
\end{align}
evaluated at the $\overline{\text{MS}}$ scale $\mu=M_S\equiv M_H\simeq M_A$. Here $0<x\equiv m_t^2/M_S^2<1$. It is found numerically that, with $M_S\simeq \mathcal{O}(1)$~TeV, the mass difference due to the top-loop correction is only of $\mathcal{O}(1)$~GeV and can be therefore neglected safely.
 
In short, the quasi-degenerate condition $M_H\simeq M_A\simeq M_{H^+}$ can be protected from large radiative corrections within the pA2HDM framework. In the next subsection, we will demonstrate that the quasi-degenerate mass spectrum is also viable to address simultaneously the $\Delta a_{e,\mu}$ anomalies in the same framework, which is realized in addition by power-enhancing the charged-lepton Yukawa couplings. At the same time, compared to the case with the linear alignment~\cite{Ilisie:2015tra,Han:2015yys,Cherchiglia:2016eui}, the slower speed of the power enhancement in the tau Yukawa coupling, together with the heavy quasi-degenerate Higgs bosons, renders the NP contributions to the $Z$- and $\tau$-decay observables smaller than the current uncertainties, and hence safely negligible.
 
\subsection{Simultaneous and correlative explanation of $\Delta a_{e,\mu}$}

In the pA2HDM, the NP contributions to $(g-2)_{e,\mu}$ consist of the one-loop diagrams mediated by $H^+$, $H$, $A$ bosons, with the corresponding amplitudes given, respectively, by
\begin{align}\label{al HA+}
\delta a_{l}^{H^+}&=-\frac{\mathcal{Y}_{\ell,l}^2}{96\pi^2}\,\frac{m_l^2}{M_{H^+}^2},
\nonumber \\[0.2cm]
\delta a_{l}^{H}&=\frac{\mathcal{Y}_{\ell,l}^2}{16\pi^2}\,\frac{m_l^2}{M_{H}^2}\,\left[\ln \frac{M_{H}^2}{m_l^2} -\frac{7}{6}\right],\nonumber \\[0.2cm]
\delta a_{l}^{A}&=\frac{\mathcal{Y}_{\ell,l}^2}{16\pi^2}\,\frac{m_l^2}{M_{A}^2}\,\left[-\ln \frac{M_{A}^2}{m_l^2} +\frac{11}{6}\right],
\end{align}
the classical two-loop Barr-Zee diagrams~\cite{Barr:1990vd,Czarnecki:1995wq}, with the dominant contributions from the top-quark and tau loops given, respectively, by
\begin{align}\label{al HAt}
\delta a_l^{H,t}&=-\frac{\alpha_{EM}}{6\pi^3}\,\frac{m_l}{m_t}\,\mathcal{Y}_{\ell,l}\, \mathcal{Y}_{u,3}\,\mathcal{F}(m_t^2/M_{H}^2),\nonumber \\[0.2cm]
\delta a_l^{A,t}&=-\frac{\alpha_{EM}}{6\pi^3}\,\frac{m_l}{m_t}\,\mathcal{Y}_{\ell,l}\, \mathcal{Y}_{u,3}\,\mathcal{G}(m_t^2/M_A^2),\nonumber \\[0.2cm]
\delta 
a_l^{H,\tau}&=-\frac{\alpha_{EM}}{8\pi^3}\,\frac{m_l}{m_\tau}\, \mathcal{Y}_{\ell,l}\,\mathcal{Y}_{\ell,3}\,\mathcal{F}(m_\tau^2/M_H^2),
\nonumber \\[0.2cm]
\delta 
a_l^{A,\tau}&=\frac{\alpha_{EM}}{8\pi^3}\,\frac{m_l}{m_\tau}\, \mathcal{Y}_{\ell,l}\,\mathcal{Y}_{\ell,3}\,\mathcal{G}(m_\tau^2/M_A^2),
\end{align}
as well as the new Barr-Zee diagram induced by the $H^+$ propagator and top-bottom quark loop~\cite{Ilisie:2015tra}, with the result given by
\begin{align}\label{al H+tb}
\delta a_l^{H^+,tb}&=\frac{3\alpha_{EM}\vert V_{tb}\vert^2}{64\pi^3 \sin\theta_W^2}\,\frac{m_l m_t}{M_{H^+}^2-M_W^2}\, \mathcal{Y}_{\ell,l}\,\mathcal{Y}_{u,3}\nonumber \\[0.2cm]
&\times \int_0^1 dx\, (x-\frac{1}{3})\,x\,(1+x)\,\left[\mathcal{Q}\left(\frac{m_t^2}{M_{H^+}^2},\frac{m_b^2}{M_{H^+}^2}\right)-\mathcal{Q}\left(\frac{m_t^2}{M_W^2},\frac{m_b^2}{M_W^2}\right)\right].
\end{align}
In the above expressions, $m_l$ is the charged-lepton mass, $\alpha_{EM}$ the electromagnetic fine-structure constant, $V_{tb}$ the CKM matrix element, and $\theta_W$ the weak mixing angle. The scalar functions $\mathcal{F}$, $\mathcal{G}$ and $\mathcal{Q}$ are defined, respectively, by
\begin{align}
\mathcal{F}(x)&=\frac{x}{2}\int_0^1 dy\,\frac{1-2y (1-y)}{y(1-y)-x}\,\ln\frac{y(1-y)}{x},
\nonumber \\[0.2cm]
\mathcal{G}(x)&=\frac{x}{2}\int_0^1 dy\, \frac{1}{y(1-y)-x}\,\ln\frac{y(1-y)}{x},
\nonumber \\[0.2cm]
\mathcal{Q}(a,b)&=\frac{1}{x(1-x)-a x-b (1-x)}\,\ln\left[ \frac{ax+b(1-x)}{x(1-x)}\right].
\end{align}
It should be noted that, even with an exact degenerate Higgs mass spectrum, the cancellation among the $H$, $A$ and $H^+$ contributions is still not exact, and thus there is a net effect on $\Delta a_{e,\mu}$. The NP effects on $\Delta a_{e,\mu}$ depend on both the flavor-universal power $n_\ell$ and the degenerate Higgs mass $M_S$, while the dependence on the quark power $n_u$ is quite insensitive, because the involved top-quark Yukawa coupling is given by $\vert\mathcal{Y}_{u,3}\vert \simeq 0.99^{n_u}\simeq 1$.   

\begin{figure}[ht]
	\centering	
	\includegraphics[width=0.49\textwidth]{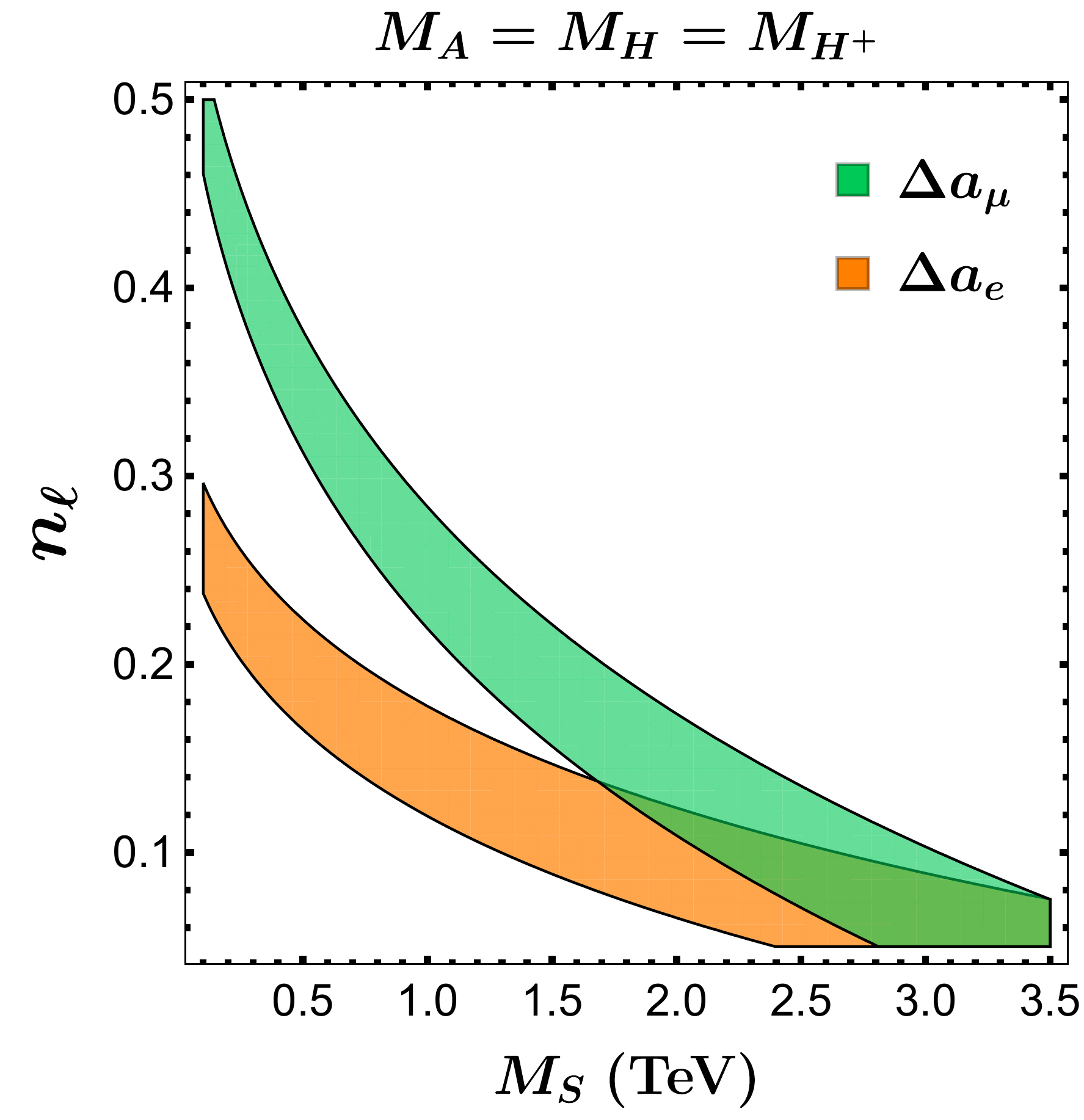}\;\,
	\includegraphics[width=0.49\textwidth]{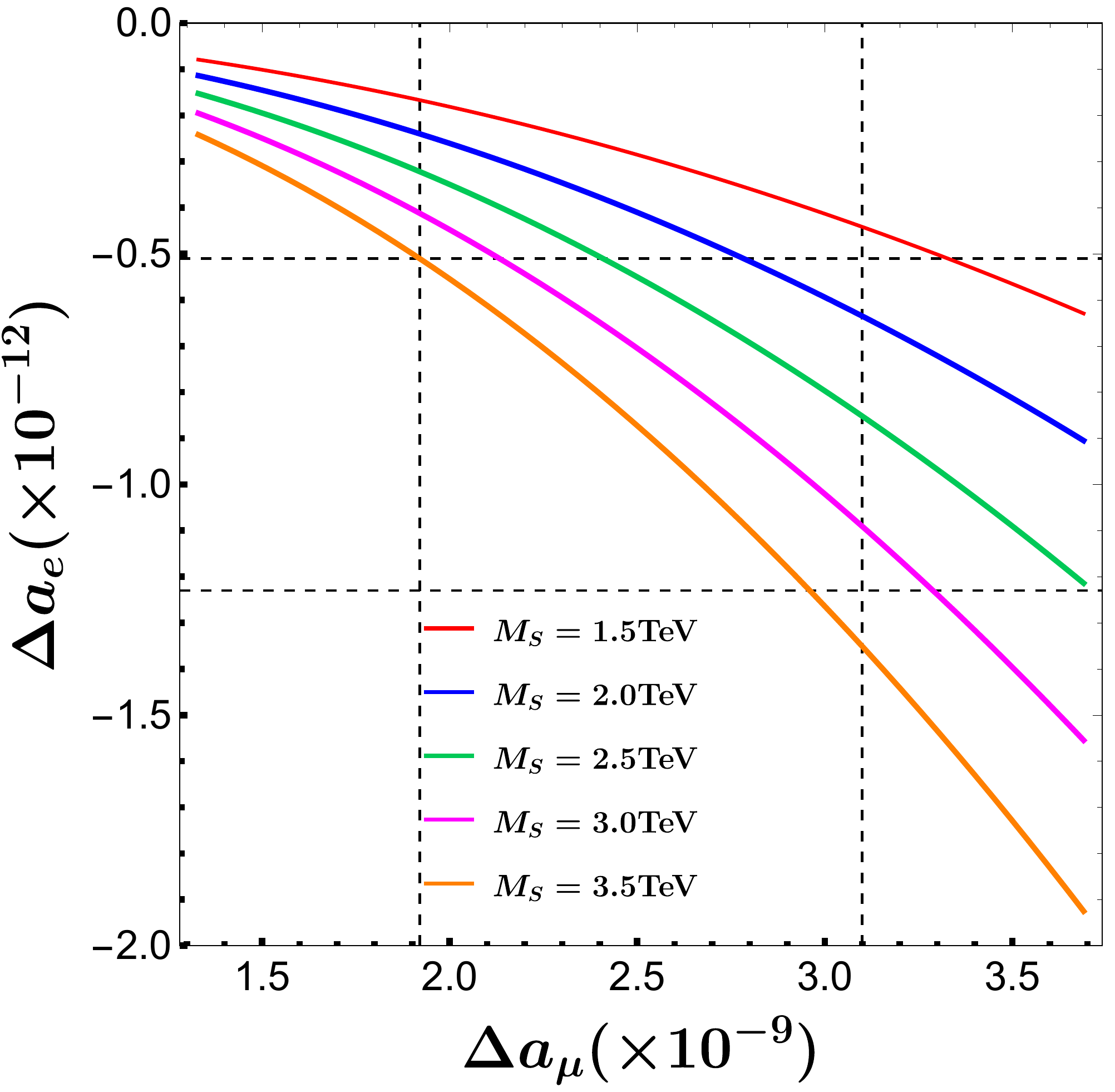}
	\caption{\label{gfit} Left: allowed parameter region in the $(M_S, n_\ell)$ plane needed to realize a simultaneous explanation of $\Delta a_{e,\mu}$ at $1\sigma$ level, in the degenerate limit of Higgs mass spectrum. Right: correlation between $\Delta a_{\mu}$ and $\Delta a_{e}$ for different choices of the degenerate Higgs mass $M_S$. The black dashed lines represent the $1\sigma$ intervals of $\Delta a_{e,\mu}$, given by eqs.~\eqref{eq:deltaamu} and \eqref{eq:deltaae}.}
\end{figure}

In the degenerate limit of Higgs mass spectrum, we depict in figure~\ref{gfit} the allowed parameter region in the $(M_S, n_\ell)$ plane (left) that can realize a simultaneous explanation of the $\Delta a_{e,\mu}$ anomalies at $1\sigma$ level, as well as the correlation between $\Delta a_{\mu}$ and $\Delta a_{e}$ for different choices of the degenerate Higgs mass $M_S$ (right). Note that, to realize such an explanation, we have postulated a positive (negative) sign in the electron (muon) entry of the NP Yukawa matrix $\mathcal{Y}_{\ell}$ by choosing $\theta_e=0$, $\theta_\mu=\pi$ and an even integer of $1/n_\ell$, which is also consistent with the required charged-lepton power $n_\ell$ shown in the left plot of figure~\ref{gfit}. For instance, the choice $n_\ell=1/10$, which corresponds to the case where $1/n_\ell$ is an even integer, allows negative entries of $\mathcal{Y}_\ell$, as discussed in section~\ref{sec:mechanism}, while a positive definite $\mathcal{Y}_u$ is always guaranteed by choosing $n_u=1$. From the left plot of figure~\ref{gfit}, we can also see that a simultaneous explanation of $\Delta a_{e,\mu}$ requires that the charged-Higgs mass $M_{H^+}\gtrsim1.7$~TeV. In addition, for even heavier Higgs bosons in the interval $[1.5, 3.5]$~TeV, the parameter region allowed is generally enlarged, which is also clear from the right plot of figure~\ref{gfit}.  

The interesting correlation between $\Delta a_{\mu}$ and $\Delta a_{e}$ shown in the right plot of figure~\ref{gfit} can be interpreted as follows. Assembling eqs.~\eqref{al HA+}--\eqref{al H+tb}, we can formally parametrize the relation between the total NP contributions to $\delta a_e$ and $\delta a_\mu$ as
\begin{align}\label{correlation g-2}
\delta a_e=a_0+a_1\delta a_\mu+a_2\delta a_\mu^2\,,
\end{align}
which results from the flavor-universal power $n_\ell$ with fixed degenerate Higgs mass $M_S$. Numerically, $a_{0}\simeq \mathcal{O}(10^{-15})$, $a_{1}\simeq \mathcal{O}(10^{-5})$, and $a_{2}\simeq \mathcal{O}(10^{4-5})$ for $M_S$ chosen in the interval $[1.5, 3.5]$~TeV. For a small range of $\Delta a_\mu$ ($\simeq \mathcal{O}(10^{-9})$), the correlation is quasi-linear, but not exactly linear, because the second and third terms in eq.~\eqref{correlation g-2} share the same order of magnitude. Such a quasi-linear tendency is also visualized in the right plot of figure~\ref{gfit}. Note that if the NP Yukawa couplings of electron and muon are entirely independent, as is the case in the gA2HDM~\cite{Jana:2020pxx,Botella:2020xzf}, there would be no correlation between $\Delta a_{\mu}$ and $\Delta a_{e}$, even when the $1\sigma$ interval of $\Delta a_{e,\mu}$ can be fitted. Thus, such a correlative explanation makes the pA2HDM distinguishable from other 2HDM candidates. 

\subsection{Compatibility with the $Z,\tau$ decay data and the LHC bounds}

Let us now consider the side effects generated by the simultaneous and correlative explanation of $\Delta a_{e,\mu}$. The additional Higgs bosons can contribute to the $Z\to \bar{\ell} \ell$ decays via the vertex corrections involving the  $H(A)$-$\ell$-$\ell$, $H(A)$-$A(H)$-$\ell$ and $H^+$-$H^-$-$\nu$ triangle loops, and the NP contributions can be encoded into the renormalized effective vertex. The associated lepton-flavor universality tests in $Z$ decays can then be parametrized as
\begin{align}\label{Z LFU}
\frac{\Gamma (Z\to \bar\ell_i \ell_i)}{\Gamma(Z\to \bar\ell_j \ell_j)}
&=R^{Z, \rm SM}_{\ell_i\ell_j}\left(1+\frac{2{\rm{Re}}\left[F_L^Z (\hat{F}_{Li}^{Z *}-\hat{F}_{Lj}^{Z*})+F_R^Z (\hat{F}_{Ri}^{Z*}-\hat{F}_{Rj}^{Z*})\right]}{\vert F_L^{Z}\vert^2+\vert F_R^{Z}\vert^2}\right),
\nonumber \\[0.2cm]
&\equiv R^{Z, \rm SM}_{\ell_i\ell_j}\,\left(1+\delta g_{Z,ij}\right),
\end{align}
where $F_L^Z=(-1/2+\sin^2\theta_W)g_2/\cos\theta_W$ and $F_R^Z=\sin^2\theta_W g_2/\cos\theta_W$ are the left- and right-handed couplings of the tree-level $Z \bar \ell \ell$ vertex, with $g_2$ being the $SU(2)_L$ gauge coupling. The one-loop renormalized effective couplings $\hat{F}_{L,R}^Z$ calculated by keeping only the logarithmic charged-lepton mass dependence are given, respectively, by
\begin{align}
\hat{F}_{L,l}^Z&=-\frac{g_2 \mathcal{Y}_{\ell,l}^2 M_Z^2}{144\pi^2 \cos\theta_W M_S^2}\,\left(3\sin^2\theta_W \ln\frac{m_l^2}{M_S^2}+4\sin^2\theta_W+1\right),
\\[0.2cm]
\hat{F}_{R,l}^Z&=\frac{g_2 \mathcal{Y}_{\ell,l}^2 M_Z^2}{576\pi^2 \cos\theta_W M_S^2}\,\left(6 \cos(2\theta_W) \ln\frac{m_l^2}{M_S^2}-14\sin^2\theta_W +13\right),
\end{align}
in the degenerate limit of Higgs mass spectrum, with $M_S\equiv M_H=M_A=M_{H^+}\simeq \mathcal{O}(1)$~TeV. It can be seen that, after normalized to the tree-level result, the one-loop vertex corrections have the size $\delta g_Z \simeq  10^{-5}\,\mathcal{Y}_{\ell,l}^2$. Given that $\mathcal{Y}_{\ell,3}\simeq \mathcal{O}(0.1)$ required by the simultaneous explanation of $\Delta a_{e,\mu}$ anomalies, the NP effect is much smaller than the current uncertainties $\mathcal{O}(10^{-4})$~\cite{ALEPH:2005ab}. By a similar argument, we have checked that the one-loop vertex corrections from $H^+$-$H^-$-$\ell$ and $H^+$-$\ell$-$\ell$ loops to the invisible decay $Z\to \bar\nu \nu$ are also negligible in the degenerate limit.

For the NP effects on the charged-lepton three-body decays, we should consider both the tree-level $H^+$-mediated contribution as well as the vertex corrections. Generically, the total invariant amplitude can be written as
$\mathcal{M}=\mathcal{M}_W+\mathcal{M}_+$, where $\mathcal{M}_W$ incorporates the tree-level and one-loop vertex corrected $W$-mediated contributions, while $\mathcal{M}_+$ denotes the tree-level $H^+$-mediated amplitude. The total amplitude squared can then be parametrized as
\begin{align}
\vert \mathcal{M}\vert^2\simeq \vert\mathcal{M}_{W0}\,\vert^2\left(1+2\text{Re}[\hat{F}^W_L/F^W_L]\right)+\vert\mathcal{M}_+\vert^2+2\text{Re}[\mathcal{M}_{W0}^*\mathcal{M}_+],
\end{align}
where the interference between loop- and tree-level NP effects has been neglected. Here $\mathcal{M}_{W0}$ denotes the tree-level $W$-mediated amplitude, and $F_L^W=g_2/\sqrt{2}$ is the coupling of the tree-level $W\ell \nu$ vertex. Note that products of the unitary PMNS matrix elements, once summed over the invisible neutrino states, do not appear in the final expression, and thus we can sufficiently use the SM $W\ell \nu$ vertex. In the degenerate limit, the vertex corrections from $H(A)$-$H^+$-$\ell$ loops cancel among themselves, leading to $\hat{F}_L^W=0$. As a consequence, the total width of a charged-lepton three-body decay can be parametrized as
\begin{align}
\Gamma(\ell_i\to\ell_j \nu \bar\nu)&=\Gamma^{\rm SM}(\ell_i\to \ell_j \nu \bar\nu)\,\left[1-2X_{ij}\,\frac{m_j\,g(m_j^2/m_i^2)}{m_i \,f(m_j^2/m_i^2)}+\frac{X_{ij}^2}{4}\right] \nonumber \\[0.2cm]
&\equiv \Gamma^{\rm SM}(\ell_i\to \ell_j \nu \bar\nu)\,\left(1-\delta g_{W,ij}\right),
\end{align}
with $X_{ij}=\mathcal{Y}_{\ell,i}\,\mathcal{Y}_{\ell,j}\,v^2/2M_{H^+}^2$. The phase-space factor, $f(x)=1-8x+8x^3-x^4-12x^2\ln x$, is known from the decay $\ell_i\to \ell_j \nu \bar\nu$ within the SM, while $g(x)=1+9x-9x^2-x^3+6x(1+x)\ln x$ stems from the interference between $W$- and $H^+$-mediated amplitudes.

Due to the cancellation of one-loop vertex corrections in the degenerate limit of Higgs mass spectrum, the lepton-flavor universality tests in $\tau$ decays receive only the tree-level $H^+$-mediated contribution, with the most significant effects arising from the ratios $g_\tau/g_e$ and $g_\mu/g_e$ via 
\begin{align}
\left(\frac{g_\tau}{g_e}\right)^2&\equiv\frac{\Gamma(\tau\to \mu \nu\bar\nu)/\Gamma^{\rm SM}(\tau\to \mu \nu\bar\nu)}{\Gamma(\mu\to e \nu\bar\nu)/\Gamma^{\rm SM}(\mu\to e \nu\bar\nu)}\simeq 1-\delta g_{W,32}+\delta g_{W,21}, \\[0.2cm]
\left(\frac{g_\mu}{g_e}\right)^2&\equiv\frac{\Gamma(\tau\to \mu \nu\bar\nu)/\Gamma^{\rm SM}(\tau\to \mu \nu\bar\nu)}{\Gamma(\tau\to e \nu\bar\nu)/\Gamma^{\rm SM}(\tau \to e \nu\bar\nu)}\simeq 1-\delta g_{W,32}+\delta g_{W,31},
\end{align}
while the effect on $g_\tau/g_\mu$ is much smaller because $\delta g_{W,32}\gg \delta g_{W,21},\delta g_{W,31}$. In addition, the NP effects on the hadronic $\tau$ decays can also be negligible, since the SM light-quark Yukawa couplings are small and the NP ones are further suppressed by the powers $n_{u,d}\geq1$. Confronted with the current results~\cite{Zyla:2020zbs,Amhis:2019ckw},
\begin{align}\label{g ratios}
\left(\frac{g_\tau}{g_e}\right)=1.0029\pm 0.0014,\quad \left(\frac{g_\mu}{g_e}\right)=1.0018\pm 0.0014, 
\end{align} 
it can be figured out that, with $n_\ell=1/10$ and $M_{H^+}=2$~TeV as a benchmark, the largest allowed size for $\delta g_{W,32}$ is given by $\delta g_{W,32}\simeq 1\times 10^{-4}$, which is compatible with that derived from eq.~\eqref{g ratios} within $2\sigma$ error bars. Clearly, the allowed value of $\delta g_W$ would become more suppressed by increasing the charged-Higgs mass.

Finally, let us discuss briefly the bounds from LHC direct searches. The constraints from B-physics observables have already pushed the charged-Higgs mass up to a few TeV. This further implies $\mathcal{O}(1)$~TeV-scale neutral Higgs bosons in light of the electroweak precision tests and a small coupling $\lambda_5$ in the scalar potential. Both the mass quasi-degeneracy and the smallness of $\lambda_5$ are actually protected by some approximate symmetries, as discussed already in subsection~\ref{Higgs spectrum}. With $\mathcal{O}(1)$~TeV Higgs bosons, the direct tests of the pA2HDM at colliders should be performed with an attention at the same scale. In ref.~\cite{Botella:2020xzf}, the new Higgs bosons with masses in the range $1-2.5$~TeV are considered to explain the $\Delta a_{e,\mu}$ anomalies in a  non-correlative manner. Furthermore, the current LHC constraints on such a mass range are also investigated in ref.~\cite{Botella:2020xzf}. It is found that, with the charged-lepton Yukawa couplings favored by the $\Delta a_{e,\mu}$ explanation, the current LHC bounds do not impose tight constraints on the Higgs masses beyond $1$~TeV. Given that the size of $\mathcal{Y}_\ell$ responsible for the $\Delta a_{e,\mu}$ explanation in the pA2HDM is comparable to that considered in the gA2HDM~\cite{Botella:2020xzf}, it is safe to expect that the current LHC bounds would not put further constraints beyond those from the B-physics observables considered here.

As a consequence, the pA2HDM proposed here is featured by a heavy quasi-degenerate Higgs mass spectrum to correlate the $\Delta a_{e,\mu}$ anomalies, and such a solution is phenomenologically safer under the $Z$ and $\tau$ decay data, as well as the current LHC bounds.

\section{Conclusion}
\label{sec:con}

In this paper, assuming that the Yukawa matrices presented in an effective 2HDM are Hermitian, which can be enforced by some more fundamental flavor and/or gauge symmetries, we have presented a power-aligned 2HDM, in which the allowed unitary flavor transformation groups are reduced from $SU(3)_{Q}\times SU(3)_{u}\times SU(3)_{d}$ in the quark and $SU(3)_{E}\times SU(3)_{\ell}\times SU(3)_\nu$ in the lepton sector to $SU(3)_{q}\times SU(3)_l$. In constructing the $SU(3)_{q}\times SU(3)_l$-invariant Yukawa interactions, the Hermiticity of the Yukawa matrices renders a power alignment between the Yukawa couplings of the two scalar doublets, which is also in line with the spirit of MFV hypothesis. 

Within such a power-aligned 2HDM framework, we have analyzed the NP effects on the $\Delta a_{e,\mu}$ anomalies. It is found that, due to the severe constraints from the mass differences $\Delta M_{d,s}$, the lower bound on the charged-Higgs mass is now pushed up to around $2$~TeV. Together with a symmetry-protected Higgs mass spectrum, it is then found that a simultaneous explanation of the $\Delta a_{e,\mu}$ anomalies can be reached with TeV-scale quasi-degenerate Higgs masses. Such a explanation is also phenomenologically safer under the $Z$ and $\tau$ decay data, as well as the current LHC bounds. Furthermore, the flavor-universal power that enhances the charged-lepton Yukawa couplings brings about an interesting correlation between the two anomalies, which makes the model different and distinguishable from other 2HDM candidates and hence testable by future precise measurements.

\section*{Acknowledgements}
This work is supported by the National Natural Science Foundation of China under Grant Nos.~12075097, 11675061, 11775092 and 11947131, as well as by the Fundamental Research Funds for the Central Universities under Grant Nos.~CCNU20TS007 and 2019YBZZ079. X Zhang is also supported by the CCNU-QLPL Innovation Fund (QLPL2019P01).

\bibliographystyle{JHEP}
\bibliography{reference}
 
\end{document}